\documentclass[twocolumn,pra,aps,showpacs]{revtex4}

\usepackage{amsmath}
\usepackage{amssymb}
\usepackage[dvips]{graphicx}
\usepackage{psfrag}
\usepackage{wasysym}

 \newcommand{\Z}{\mathbf{Z}}

 \newcommand{\sset}[1]{ \{#1\} }

 \newcommand{\prima}{^\prime}

 \newcommand{\bbar}[1]{\bar{\bar {#1}}}

 \newcommand{\ket}[1]{|#1\rangle}
 \newcommand{\bra}[1]{\langle #1|}

 \newcommand{\mr}{\mathrm{r}}
 \newcommand{\mg}{\mathrm{g}}
 \newcommand{\mb}{\mathrm{b}}
 \newcommand{\bc}{{\bar{c}}}
 \newcommand{\bbc}{{\bbar{c}}}
 \newcommand{\HilbColor}{\mathcal H_{\mathrm C}}

 \newcommand{\rel}{,}
\begin{document}

\title[Short Title]{
Interacting Anyonic Fermions in a Two-Body Color Code Model}

\author{H. Bombin$^{1}$, M. Kargarian$^{2}$ and M.A. Martin-Delgado$^{3}$}
\affiliation{ $^1$Department of Physics, Massachusetts Institute of Technology, 
Cambridge, Massachusetts 02139, USA \\
$^2$Physics Department, Sharif University of Technology, Tehran 11155-9161,
Iran\\$^3$Departamento de F\'{i}sica Te\'orica I,
Universidad Complutense, 28040 Madrid, Spain}

\begin{abstract}
We introduce a two-body quantum Hamiltonian model of spin-1/2 on a 2D spatial lattice with exact topological degeneracy in all coupling regimes.
There exists a gapped phase in which the low-energy sector reproduces an effective color code model. High energy excitations fall into three families of anyonic fermions that turn out to be  strongly interacting. The model  exhibits a  $\Z_2\times \Z_2$ gauge group symmetry and string-net integrals of motion,  which are related to the  existence of topological charges that are invisible to moving high-energy fermions.
\end{abstract}



\maketitle

\section{Introduction}
\label{sect_I}

The Kitaev model on the honeycomb lattice \cite{honeycomb} has attracted a
great deal of attention in the recent years
\cite{Feng_et_al07,Yao_Kivelson07,Yang_et_al07, fermions_disturb} since it offers
the opportunity to study many properties of topologically ordered
systems in a very well-suited scenario for condensed matter.
One of the most interesting properties of the Kitaev model is that one of its phases effectively reproduces the
famous toric code. This is the first example of topological quantum error correction and plays a major role in quantum
information.

Yet, there is another family of topological codes, 
color codes,
which exhibit very remarkable properties. They allow an appropriate implementation of the whole Clifford group of quantum gates,
which are essential in quantum information tasks\cite{topologicalClifford}.
Topological color codes (TCC) are constructed with quantum lattice
Hamiltonians that typically demand 6-body terms
in a 2D spatial lattice, requirements quite unrealistic in a condensed matter framework. Up to now, it has remained a challenge to find a 2-body Hamiltonian capable of hosting a TCC in a particular coupling regime.

In this paper we provide such a 2-body spin-${\frac 1 2}$ Hamiltonian model. It turns out to exhibit very rich physics and,
quite remarkably, it does not belong to the family of models originated after Kitaev's model. These latter models are defined on trivalent lattices, which allows a fermionization yielding a free fermion exact solution. Instead, the lattice of our model is 4-valent, a sharp difference that prevents complete solvability and gives rise to very interesting features not present in the mentioned models:

\noindent i/ Exact topological degeneracy in all coupling regimes, rooted on the existence of string-net integrals of motion. This degeneracy, related to certain `invisible' topological charges, is $4^g$-fold in surfaces of genus $g$.

\noindent ii/ Emergence of three families of strongly interacting fermions with semionic relative statistics.

\noindent iii/ An exact $\Z_2\times \Z_2$ gauge symmetry.
Each family of emergent fermions sees a different $\Z_2$ gauge subgroup.

\noindent Thus, the model admits an exact analysis of many interesting properties. This is so mainly because there exist local integrals of motion in an amount of $\frac 1 3$ of the total number of spins.

Although we will focus on a particular phase of the model, namely the one that effectively yields TCCs, a rich phase diagram beyond this gapped phase is to be expected in analogy with \cite{honeycomb}. This includes the possibility
of non-Abelian anyons and other phases with interesting many-body effects.
In addition, it is possible to break a symmetry of the model, which we call color symmetry, whereas the exact features above are kept. This paves the way to a yet more complex phase diagram.

The properties of the model make it a good candidate for
an experimental realization, by means
of some engineering scheme like polar molecules on optical lattices\cite{polar_molecules05}. It is also
amenable to numerical computations \cite{numerics08}
with several methods, which would help to uncover
some of its non-perturbative aspects and phases.

In summary, the model represents a relevant contribution to the
difficult task of searching for systems with emerging anyons and topological order, which are conceptually important phenomena and a source of new physics.

\begin{figure}
\psfrag{a}{a}
\psfrag{b}{b}
\psfrag{c}{c}
\psfrag{x}{\tiny $x$}
\psfrag{y}{\tiny $y$}
\psfrag{z}{\tiny $z$}
\psfrag{mr}{\tiny $r$}
\psfrag{mg}{\tiny $g$}
\psfrag{mb}{\tiny $b$}
 \includegraphics[width=8.8cm]{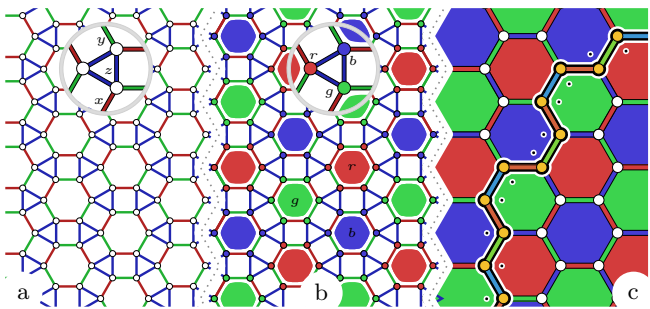}
\caption{Three different points of view of the system. (a) is the physical one, with vertices representing spins and colored links representing $\sigma^x_a\sigma^x_b$, $\sigma^y_a\sigma^y_b$ and $\sigma^z_a\sigma^z_b$ two-body interactions. In (b) both spins and hexagonal plaquettes have been colored. (c) is the reduced lattice $\Lambda$, where vertices represent sites, that is, the blue triangles of (a).
We show a magnified site in (a,b) and a closed string in (c).}
\label{figure_lattice}
\end{figure}

This paper is organized as follows: in Sect. \ref{sect_II} we 
introduce the quantum Hamiltonian model based solely on 2-body interactions
between spin-$\frac{1}{2}$ particles. The lattice is two-dimensional and has
coordination number 4, instead of the usual 3 for the Kitaev model.
It is pictured in Fig.~\ref{figure_lattice} and it is called ruby lattice.
In Sect. \ref{sect_III}, we introduce a mapping from the original 
spin-$\frac{1}{2}$ degrees of freedom onto bosonic degrees of freedom
in the form of hard-core bosons which also carry a pseudospin.
In Sect. \ref{sect_IV}, we describe the constants of motion of our model,
specifying both its various types (normal strings and non-standard stringnets),
as well as its number of $\frac 1 3$ of the total number of spins.
With the help of the previous mapping, it is possible to analyse very
interesting qualitative features of our model at non-perturbative level.
In Sect. \ref{sect_V}, we show how the emerging quasiparticle excitations
of our model are anyonic fermions with strong interactions, which are
a manifestation of the string-net structure of the constants of motion.
Thus, the model is not a free fermion model.
In Sect. \ref{sect_VI}, we solve the problem of finding a 2-body 
quantum lattice Hamiltonian hosting the topological color code
as one of its phases. Interestingly enough, the relevant properties
of the TCC remain valid at the non-perturbative level as well.
We also describe the notion of invisible charges
and discuss anyon condensation, and its implications.
Sect. \ref{sect_conclusions} is devoted to conclusions.


\section{Hamiltonian Model with 2-Body Interactions}
\label{sect_II}

The model of interest lives in the lattice of Fig.~\ref{figure_lattice}(a). Notice that, unlike in
\cite{honeycomb,Feng_et_al07,Yao_Kivelson07,Yang_et_al07}, it has
coordination four. Vertices represent spin-${\frac 1 2}$ systems and links two-body interactions. The Hamiltonian is
\begin{equation}\label{Hamiltonian_A}
 H = -\sum_{<i, j>} J_w \,\sigma^w_i \sigma^w_j,\qquad w=\begin{cases}x,&\text{red links}\\y,&\text{green links}\\z,&\text{blue links}\end{cases}
\end{equation}
where the couplings $J_w\neq 0$ are real, $\sigma^w$ are the Pauli matrices and the sum extends over all links. Notice that blue links are special since they form triangles, which we will refer to as {\em sites} for reasons that will be apparent below.

Sites are the vertices of a reduced lattice $\Lambda$ in which links are given by the pairs of parallel red and green links of the original lattice, see Fig.~\ref{figure_lattice}(c). This reduced lattice will play an essential role in the study of the properties of the model. It is an hexagonal lattice and thus has 3-colorable plaquettes. We choose to color them with red ($\mr$), green ($\mg$) and blue ($\mb$). We also color links accordingly, in such a way that $c$-links connect $c$-plaquettes. Here and in what follows, we use the letter $c$ to denote color variables. We remark that this coloring has nothing to do with the one used for links in Fig.~\ref{figure_lattice}(a). The coloring of the reduced lattice induces a coloring of some plaquettes of the original lattice, see Fig.~\ref{figure_lattice}(b), and also a vertex coloring that labels the spins within a site.

Although we have given a particular lattice for concreteness, the analysis that follows is much more general. Instead of having a hexagonal reduced lattice, it could be any trivalent lattice with 3-colorable plaquettes. Such lattices were named 2-colexes in \cite{branyons}. Any closed surface will work, orientable or not. That the construction works on non-orientable surface is a consequence of the equivalence of vertex bicolorability and orientability in 2-colexes \cite{twoBodyColor}.


\section{Bosonic Mapping}
\label{sect_III}

In section \ref{sect_VI} we will analyze in detail the regime $J_z
>0$, $|J_x|,|J_y|\ll J_z$. Let us set for simplicity $J_z=1/4$.
Then in the extreme case $J_x=J_y=0$ the system consists of isolated triangles, one per
site. In an energy eigenstate, each of them contributes an energy $-3/4$ or $1/4$, so
that we can attach a quasiparticle with energy gap equal to 1 to
each triangle. With this motivation, we give here a map to a new system in which these quasiparticles are explicit. The mapping
is \textit{exact}, so that only the physical picture is changed. It has the advantage of isolating those degrees of freedom that survive once the integrals of motion to be described below have been fixed.

In Fig.~\ref{figure_lattice}~(b) each spin in a site has been identified with a color. We label the
corresponding Pauli operators as $\sigma^w_c$ with $c=\mr,\mg,\mb$.
Consider the Hilbert space $\HilbColor$ with orthonormal
basis $\sset{\ket 0, \ket \mr,\ket \mg,\ket \mb}$ and introduce the
colored annihilation operators 
\begin{equation}\label{mapping}
b_c := \ket 0 \bra c, \qquad c=\mr,\mg,\mb,
\end{equation}
so that $\HilbColor$ represents a hardcore boson with three possible
color states. The number operator $n$ and the colored number operator $n_c$ are
\begin{equation}\label{mapping}
n :=\sum_c n_c, \qquad n_c :=
b^\dagger_c b_c.
\end{equation}
At each site, we attach such a boson and also an
effective spin-${\frac 1 2}$. We have to relate this degrees of freedom to the original three spins in the site, which are colored as in Fig.~\ref{figure_lattice}(b). The mapping can be expressed by relating bases of both systems. In particular, taking the usual up/down
basis for the three physical spins
and the tensor product basis 
\begin{equation}
\ket {a, d}=\ket a \otimes \ket d, \qquad a=\uparrow,\downarrow, \quad d=0,\mr,\mg,\mb, 
\end{equation}
for the effective spin and hardcore boson system we have
\begin{alignat}{4}\label{mapping_basis}
&\ket {\uparrow, 0} &\equiv \ket{\uparrow  \uparrow \uparrow},\quad
&\ket {\downarrow, 0} &\equiv \ket{\downarrow  \downarrow \downarrow},\\
&\ket {\uparrow, \mr} &\equiv \ket{\uparrow  \downarrow \downarrow},\quad
&\ket {\downarrow, \mr} &\equiv \ket{\downarrow  \uparrow \uparrow},\\
&\ket {\uparrow, \mg} &\equiv \ket{\downarrow \uparrow \downarrow},\quad
&\ket {\downarrow, \mg} &\equiv \ket{\uparrow \downarrow \uparrow},\\
&\ket {\uparrow, \mb} &\equiv \ket{ \downarrow \downarrow \uparrow },\quad
&\ket {\downarrow, \mb} &\equiv \ket{ \uparrow \uparrow \downarrow }.
\end{alignat}
More compactly, the mapping can be expressed by identifying operators as follows
\begin{equation}\label{mapping}
\sigma^z_c \equiv \tau^z \otimes p_c, \quad \sigma^v_c \equiv \tau^v \otimes (b_c^\dagger + b_c + s_v r_c),
\end{equation}
where $v=x,y$, $s_x:=-s_y:=1$, the symbols $\tau$ denote the Pauli operators on the effective
spin and we are using the color parity operators $p_c$ and the color switching operators $r_c$ defined as
 \begin{equation}
 p_c := 1-2(n_{\bar c}+n_{\bbar c}),\qquad r_c := b^\dagger_\bc b_\bbc + b^\dagger_\bbc b_\bc,
 \end{equation}
where the bar operator transforms colors cyclically as follows: 
\begin{equation}
\bar \mr := \mg,\quad \bar \mg := \mb, \quad\bar \mb := \mr.
\end{equation}

\begin{figure}
\psfrag{a}{a}
\psfrag{b}{b}
\psfrag{c}{c}
\psfrag{d}{d}
\psfrag{1}{\tiny 1}
\psfrag{2}{\tiny 2}
\psfrag{3}{\tiny 3}
 \includegraphics[width=8.8cm]{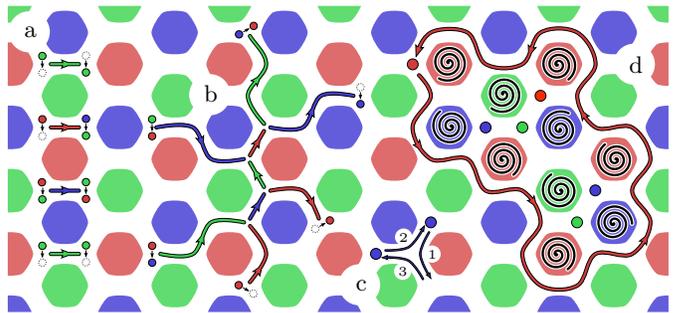}
\caption{Several $c$-fermion processes, in the effective hexagonal lattice. $c$-fermions are depicted as $c$-colored balls, and their movement with $c$-colored directed strings. When the initial and final color charge at a site are different before and after a process, we show the transition with an small black arrow. (a) Examples of the effects of the terms in \eqref{Hamiltonian_B}. From top to bottom, the corresponding terms are $t_\mg^\mr$, $u_\mr^\mb$, $r_\mb^\mg$ and $v_\mg^\mr$. (b) A typical $c$-fermion interaction process, with hoppings, fusions and splittings. (c) The exchange of two identical $\mb$-fermions. (d) An $\mr$-fermion surrounds a region $\lambda$. The phase that it picks up depends both on the number of $\mg$- and $\mb$-fermions in $\lambda$ and the vortex states of the plaquettes marked with an spiral.}\label{figure_fermions}
\end{figure}

From this point on we will be working always in the reduced lattice $\Lambda$. For compactness, we will use
a simplified notation in which site indices are supressed and
only relative positions are indicated: the notation $O_{\rel c}$
means $O$ applied at the site that is connected to a site of reference by a
$c$-link. For example, if $i,j,k$ are neighboring sites, with $i$ the reference site and $j, k$ are connected to $i$ by a $\mg$-link and a $\mr$-link respectively, then instead of $A_iB_jC_k$ we write $AB_{,\mg}C_{,\mr}$. That is, we only indicate the relationship between sites, and completely omit the reference site $i$, which is therefore implicit. Also, we indicate the $w=x,y,z$ indices of $J_w$, $\tau^w$, $s_w$ in terms of two colors as follows: 
\begin{equation}
c|c:=z,\quad \bc|c:=x, \quad \bbc|c:=y.
\end{equation}
Then, the Hamiltonian \eqref{Hamiltonian_A} can be exactly transformed into a new form 
\begin{equation}\label{Hamiltonian_B}
H=-3N/4+Q-\sum_{\Lambda}\sum_{c\neq c\prima} J_{c\prima|c}\, T_c^{c\prima}, 
\end{equation}
with $N$ the number of sites, $Q:=\sum_\Lambda n$ the total number of hardcore bosons,  the first sum running over the $N$ sites of the reduced lattice, the second sum running over the 6 combinations of different colors $c,c\prima$ and\begin{equation}\label{Terms}
 T_c^{c\prima} = u_c^{c\prima}+ \frac {t_c^{c\prima}+v_c^{c\prima}} 2 + \frac {r_c^{c\prima}} 4 + \mathrm{h.c.},
\end{equation}
a sum of several terms for an implicit reference site, according to the notation convention we are using. The meaning of the different terms in \eqref{Terms} is the following. The operator $t_c^{c\prima}$ is a $c$-boson hopping, $r_c^{c\prima}$ switches the color of two $\bc$- or $\bbc$-bosons, $u_c^{c\prima} $ fuses a $c$-boson with a $\bc$-boson (or a $\bbc$-boson) to give a
$\bbc$-boson ($\bc$-boson) and $v_c^{c\prima}$ destroys a pair of
$c$-bosons. A pictorial representation is offered in Fig.~\ref{figure_fermions}(a). The explicit expressions are 
\begin{alignat}{2}\label{terms}
 &t_c^{c\prima} := \tau_c^{c\prima} b_c b^\dagger_{c\rel
c\prima},\qquad &r_c^{c\prima} := \tau_c^{c\prima} r_c r_{c\rel
c\prima},\\ &u_c^{c\prima} :=  s_{c\prima|c} \tau_c^{c\prima} b_c r_{c\rel c\prima}, \qquad &v_c^{c\prima} := \tau_c^{c\prima} b_c b_{c\rel
c\prima},
\end{alignat} 
where we are using the notation  
\begin{equation}
\tau_c^{c\prima}:=\tau^{c\prima|c}\tau_{\rel c\prima}^{c\prima|c}.
\end{equation} 


\section{Constants of Motion}
\label{sect_IV}

In this section we explore several integrals of motion of the Hamiltonian \eqref{Hamiltonian_B}.

\subsection{Color charge}

A basic property of the Hamiltonian \eqref{Hamiltonian_B} is the existence of a
$\Z_2\times \Z_2$ charge. Let us label the elements of
$\Z_2\times\Z_2$ with colors as $\sset{e, \mr, \mg, \mb}$, where $e$ is the
identity element. We also label its irreps as $\sset{\chi_e, \chi_\mr,\chi_\mg, \chi_\mb}$ setting $\chi_e(c):=\chi_c (c) :=-\chi_c(\bc):=1$. Let us arbitrarily attach a charge to each family of hardcore bosons. In particular, we attach the irrep $\chi_c$ to each $c$-boson. Let $Q_c=\sum_\Lambda n_c$ be the total number of
$c$-bosons. The total color charge 
\begin{equation}\label{color_charge}
\chi_\Lambda:= \chi_\mr^{Q_\mr}\chi_\mg^{Q_\mg}\chi_\mb^{Q_\mb}
\end{equation}
is preserved by the Hamiltonian. This can be checked directly or noting that the operators
\begin{equation}
\chi_\Lambda(c) = (-1)^{Q_\bc+Q_\bbc}=\prod_\Lambda p_c
\end{equation}
commute with all the terms in \eqref{Terms}. In section \ref{sect_V} the meaning of color charge will become clear, when we relate it to a $\Z_2\times\Z_2$ gauge field.

\subsection{Plaquette operators}

We describe now certain plaquette operators that commute
with each other and with each of the terms of the Hamiltonian. As we will show below, the corresponding degrees of freedom should be regarded as vortices. For each color $c$ and each plaquette $\pi$ there is a constant of motion of the form 
\begin{equation}
S_\pi^c := \prod_{\pi} \tau^{c|c\prima} p_{c\prima\star c},
\end{equation}
where $c\prima$ is the color of the plaquette $\pi$, the product runs through its sites and $\star$ is just a convenient symmetric color operator defined by 
\begin{equation}
c\star c:=c,\qquad c\star \bc:= \bc\star c:=\bbc.
\end{equation}
This operators are not all independent. They are subject only to the following constraints
\begin{equation}\label{constraints}
\prod_{\pi\in\Lambda} S_{\pi}^c = (-1)^{N/2}, \qquad \prod_{c=\mr, \mg,\mb} S_{\pi}^c = (-1)^{s/2},
\end{equation}
where $s$ is the number of sites of a given plaquette $\pi$, and we are supposing in the first equation that the lattice forms a closed surface. It follows that there are $2F-2$ independent plaquette operators, with $F$ the number of faces, or plaquettes, of the reduced lattice $\Lambda$. The total color charge is not independent of plaquette operators, because
\begin{equation}\label{charge_constraints}
\chi_\Lambda(c)=\prod_{\pi\in \Lambda_{\bc}} S_{\pi}^{\bc}\prod_{\pi\in \Lambda_{\bbc}} S_{\pi}^{\bbc},
\end{equation}
where $\Lambda_c$ denotes the subset of $c$-plaquettes of $\Lambda$.

In section \ref{sect_V} we show that plaquette degrees of freedom can be regarded as vortices, as they correspond to a $\Z_2\times\Z_2$ gauge field, which has therefore no dynamics. The correspondence between plaquette operator eigenstates and group values is as follows.  First, for each $c$-plaquette we introduce the following alternative plaquette operators:
 \begin{align}\label{Bs}
B_\pi^\bbc &:=j_x^{s/2}\, S_\pi^\bc,\nonumber\\
B_\pi^\bc &:=j_y^{s/2} \,S_\pi^\bbc,\nonumber\\
B_\pi^c &:=(-j_xj_y)^{s/2}\,  S_\pi^c,
\end{align}
where $j_w:=J_w/|J_w|$. 
The element $g_\pi\in\Z_2\times\Z_2$ that we attach to each plaquete $\pi$ is determined, for a given eigenstate of the operators \eqref{Bs}, by the conditions
\begin{align}\label{gpi}
\chi_c(g_\pi) = B_\pi^c.
\end{align}
These equation always have a solution because $(B_\pi^c)^2 = B_\pi^\mr B_\pi^\mg B_\pi^\mb  = 1$. Given a region or collection of plaquettes $\lambda$, we will use the notation
$g_\lambda:=\prod_{\pi\in\lambda} g_\pi$.

\subsection{String operators}

Plaquette constants of motion can be generalized to closed strings and, for that matter, also to string-nets\cite{twoBodyColor}. For a closed string we mean a connected path in the lattice with no endpoints, see Fig.~\ref{figure_lattice}(c). To any such string $\gamma$ and color $c$ we attach a string operator 
\begin{equation}
S^c_\gamma:=\prod_\gamma \tau^{c|c\prima}p_{c\star c\prima},
\end{equation}
where the product runs through
the sites of $\gamma$ and, for each site, $c\prima$ is the color of the
plaquette arround which $\gamma$ turns at the site. This is exemplified in Fig.~\ref{figure_lattice}(c), where we have marked with a dot the corresponding plaquette for each vertex of the string. As in the case of plaquette operators, we have $(S_\gamma^c)^2=(-1)^{\frac
s 2} S_\gamma^c S^\bc S_\gamma^\bbc=1$ with $s$ the number of sites of $\gamma$.
Closed string operators commute with all plaquette operators and Hamiltonian terms, but not always with each
other. Namely, if the strings $\gamma$, $\gamma\prima$ cross once then $[S_\gamma^c,S_{\gamma\prima}^c]=0$ but $\sset {S_\gamma^c,S_{\gamma\prima}^\bc}=0$.

This anticommutation property, which remarkably is not present in \cite{honeycomb}, turns out to be crucial. In systems with nontrivial topology it is the source of an exact degeneracy of the Hamiltonian. For example, the lattices of Fig.~\ref{figure_lattice} with periodic boundary conditions live on a torus, which has two independent nontrivial loops $\gamma$, $\gamma\prima$. Then $S_\gamma^c$ anticommutes with $S_{\gamma\prima}^\bc$ and commutes with $S_{\gamma}^\bc$ and $S_{\gamma\prima}^c$, showing\cite{Oshikawa} that the Hamiltonian is at least 4-fold degenerate. More generally, in a closed surface with Euler characteristic $\chi$ the degeneracy is $2^{2-\chi}$. In particular, in orientable surfaces of genus $g$ there exist a $4^g$-fold topological degeneracy. This result is exact and independent of the particular phase of the system we are in. In order to label the global flux degrees of freedom attached to nontrivial string operators we can use the eigenvalues of $2-\chi$ topologically nontrivial and independent closed string operators of a given color.


\section{Anyonic Fermions}
\label{sect_V}

In the previous section we have learned that the system is divided into sectors with a given vortex and global flux configuration. We now want to understand the nature of the degrees of freedom in each of these sectors, and we start counting them. In terms of spin-1/2 degrees of freedom, the original system has $3N$ spins, the plaquette operators remove $2F-2$ of them and the nontrivial string operators of a given color $2-\chi$ more. These leaves $N\prima=2N-\chi$ effective spins in a given sector. Since we are not interested in global degrees of freedom, let us suppose that the topology is trivial, so that $2-\chi=0$. Then the resulting value of $N\prima$ can be easily understood: we are only left with hardcore boson degrees of freedom (hence the $2N$), subject to the constraint of color charge preservation (hence the $-2$). In fact, in each sector we can choose a basis with elements labeled by the state of the hardcore bosons, so that these become the relevant degrees of freedom. As we will see, the dynamics of the system transform the hard-core bosons into three families of fermions with anyonic statistics between them. Moreover, we will see that these fermions interact strongly. For arbitrary couplings this only provides us with a picture to understand the system, since the emergent quasiparticles could be very different. However, in the regime analyzed in section \ref{sect_VI} the anyonic fermions come to life explicitly.

In order to check the statistics of the effective quasiparticles emerging from hardcore boson degrees of freedom it suffices\cite{WenFermions} to study the hopping terms  in \eqref{Hamiltonian_B}. Strictly speaking, the situation here is not the same as in \cite{WenFermions}, as we are not studying quasiparticles. More close is the approach in \cite{honeycomb}, where the emphasis is done on the properties of string operators. Now,we first note that the hopping terms $t_c^\bc$ and $t_c^{\bbc}$ that appear in \eqref{Hamiltonian_B} are only enough to hop a $c$-boson around a $c$-plaquette. We need also composite hoppings: $t_c^c=u_\bbc^c {u_{\bc,c}^{c\,\dagger}}=u_\bc^c {u_{\bbc,c}^{c\,\dagger}}$ hops a $c$-boson from a $c$-plaquette to another. Notice that this notation completely agrees with \eqref{terms}. Consider a state with only two $c$-boson excitations, located at two sites separated respectively by a $\bc$ and a $\bbc$ link from a given reference site, as in Fig.~\ref{figure_fermions}(c). We may then consider a hopping process, indicated with numbers in the figure, in which the $c$-bosons are exchanged in such a way that local contributions cancel \cite{honeycomb}. The net effect of exchanging the $c$-bosons is
\begin{equation}\label{fermions}
 t_{c}^\bbc \,t_{c,c}^c \, t_{c}^\bc \,  t_{c,\bbc}^\bbc \, t_c^c \, t_{c,\bc}^\bc   =
(\tau^y\tau^y_{,\bbc}\tau^z\tau^z_{,c}\tau^x\tau^x_{,\bc})^2=-1,
\end{equation}
showing that the hardcore-bosons give rise to fermions\cite{WenFermions}.

These fermions carry a nontrivial color charge, and thus we have
three families of them. They are not free but interacting, as
follows from the existence of the $u_c^{c\prima}$ terms
in the Hamiltonian, which correspond to a 3-fermion
interaction vertex, see Fig.~\ref{figure_fermions}(b). In fact, the existence of this vertex indicates that we are not dealing just with fermions, but rather with anyons: three conventional fermions cannot fuse into the vacuum. In addition,
fermions must be coupled to a nontrivial gauge field\cite{WenFermions}.

In order to understand these two issues, let
us consider a process in which a $c$-fermion is carried around a
region $\lambda$, as in Fig~\ref{figure_fermions}(d). For
clarity, we assume that no fermions but that to be transported are
present along the boundary of $\lambda$.
The hopping process yields a phase
\begin{equation}\label{phase}
\phi^c_\lambda=\chi_c (g_\lambda) \, (-1)^{n_\bc^\lambda+n_\bbc^\lambda+n_4^\lambda}
\end{equation}
where $n_c^\lambda$ denotes the number of $c$-fermions inside
$\lambda$ and $n_4^\lambda$ the number of 2-colex plaquettes inside $\lambda$ with a number of edges that is a multiple of four. 
We conclude that each family of fermions carries a different representation of a $\Z_2\times\Z_2$ gauge group with values $g_\lambda$ dictated by the vortex states. 
In addition, fermions of different color have semionic mutual statistics: they pick a -1 phase when one of them winds around the other. Thus, fermions not only interact via virtual
fermion exchanges, but also topologically.


\section{Effective Topological Color Code}
\label{sect_VI}

We now turn to a perturbative study of the regime $J_z>0$, $|J_x|,|J_y|\ll J_z$, for which the previous bosonic mapping is specially suited. As in \cite{honeycomb_bosonization}, we apply the PCUTs method \cite{PCUTS} (perturbative continuous unitary transformations). This method produces an effective Hamiltonian $H_\mathrm{eff}$, at a given perturbation order, such that $[H_{\mathrm{eff}},Q]=0$. We are specially interested in the low-energy physics, that is, the $Q=0$ sector of the Hamiltonian (notice that the sectors here have nothing to do with the ones discussed in the previous section!). For each $c$-plaquette $\pi$ let us set $B_\pi^x:=B_\pi^{\bbc}$, $B_\pi^y:=B_\pi^{\bc}$. Then since $Q=0$ we have 
\begin{equation}
B_\pi^x=j_x^{s/2}\prod_\pi \tau^x,\qquad B_\pi^y=j_y^{s/2}\prod_\pi \tau^y.
\end{equation}
For the particular lattice of Fig.~\ref{figure_lattice}, the 9th order perturbative calculation yields, up to a constant, the effective Hamiltonian
\begin{equation}\label{Hamiltonian_eff_0}
 H_{\mathrm{eff}} = - \sum_{\pi\in\Lambda} \left ( k_x B^x_\pi + k_yB^y_\pi + k_z B^x_\pi B^y_\pi \right),
\end{equation}
with
\begin{align}\label{coeff}
k_z\!&=\!\frac {3 } 8 |J_xJ_y|^3\!+\!O(J^7),\\ 
\frac{k_x}{|J_y|^3}\!&=\!\frac{k_y}{|J_x|^3}\!=\!\frac {55489}{13824}{|J_xJ_y|^3}.
\end{align}
The Hamiltonian \eqref{Hamiltonian_eff_0} describes a color code model on the effective spins\cite{topologicalClifford}. Its ground state is the vortex free sector, where $\chi_c(g_\pi)=1$ for any plaquette $\pi$. Excitations are vortices, gapped and localized at plaquettes with $ B^x_\pi=-1$ and/or $B^y_\pi=-1$. At higher orders in perturbation theory the terms of the effective Hamiltonian take the form of products of vortex operators, which gives rise to vortex interactions\cite{twoBodyColor, honeycomb_bosonization}.

Color code models have a topological order described by a $\Z_2\times \Z_2$ quantum double\cite{quantumDouble}. Thus there exist 16 topological charges in the model, labeled with pairs $(g,\chi)$ with $g\in \Z_2\times \Z_2$  and $\chi$ an irrep of this group. The trivial charge or vacuum sector is $(e,\chi_e)$, the charges $(c,\chi_e)$, $(e,\chi_c)$, $(c,\chi_c)$ are bosons and $(c,\chi_\bc)$, $(c,\chi_\bbc)$ are fermions. As for the mutual statistics, moving a $(g,\chi)$ charge around a $(g\prima, \chi\prima)$ charge gives a topological phase $\chi(g\prima)\chi\prima(g)$. Regarding fusion rules, a $(g,\chi)$ charge and a $(g\prima, \chi\prima)$ charge form together a $(gg\prima, \chi\chi\prima)$ charge. Notice how the three fermions $(c,\chi_\bc)$ form a family closed under fusion. The same is true for the three $(c,\chi_\bbc)$ fermions. Two fermions from different families have no topological interaction. This property will turn out to be important in the understanding of the model. 

There are several possible ways to label the excitations of the gapped phase \eqref{Hamiltonian_eff_0}. Our choice is that an excitation at a $c$-plaquette $\pi$ has $(c,\chi_e)$ charge if $-B^x_\pi=B^y_\pi=1$, $(e,\chi_c)$ charge if $B^x_\pi=-B^y_\pi=1$ and $(c,\chi_c)$ charge if $B^x_\pi=B^y_\pi=-1$.

\subsection{Beyond the low-energy sector}

We now turn our attention to the high-energy quasiparticles, that is, those which contribute to the quasiparticle counter $Q$. From the analysis of section \ref{sect_V} we now that these are three families of interacting anyonic fermions. The virtual processes of these fermions create the gap for low-energy excitations, as well as interactions between them.

High-energy quasiparticles belong to a particular topological charge superselection sector. This correspondence between low- and high-energy excitations follows from the global constraints \eqref{constraints}, \eqref{charge_constraints}, which have local consequences. Namely, the local creation of a single $c$-fermion must be accompanied by the change of sign of several vortex operators, with the net result of the creation/annihilation of a low-energy $(\bc,\chi_\bbc)$ charge. This topological charge has exactly the properties needed to agree with the fermion fusion rules and the results \eqref{fermions} and \eqref{phase} derived before. 

Let us now take a closer look to the sector with $Q=1$, which involves no interactions. Using the PCUTs method one can derive the effective Hamiltonian for the gapped phase in the sector with a single $c$-fermion, for a given color $c$. At first order the $c$-fermion just hops around $c$-plaquettes, so that for $J=J_x=J_y$ we get a $-2J$ contribution to the energy gap coming from this orbital motion. Going to second order we get a non-flat dispersion relation that corresponds to the triangular lattice formed by $c$-plaquettes. The gap, at this order, is given by $1-2J-J^2/2$ and thus it closes at $J\simeq 0.45$. This is just an approximate estimation, since we are omitting all fermion interactions and, perhaps more importantly, we are taking $J\simeq J_z$. However, it is to be expected that as the couplings $J_x\sim J_y$ grow in magnitude the gap for high-energy fermions will reduce, producing a phase transition when the gap closes. Such a phase transition resembles the anyon condensations discussed in \cite{condensation}, \cite{bais_joost}. Here we are dealing with a condensation of three anyonic fermions that does not fit those examples, but part of the physical picture could be similar. There exist three nontrivial topological charges that do not interact with the three condensed anyons. We would thus expect a residual topological order in the new phase, with three nontrivial topological charges with relative semionic statistics. These charges would then be responsible for the $4^g$ exact topological degeneracy. This conjecture is to some extent supported by the validity in all phases of closed string integrals of motion which, as we will see next, represent processes related to the three topological charges invisible to the moving fermions.

\subsection {Invisible topological charges} 

It is natural to expect that open $c$-string operators will create, transport and destroy a particular topological charge among the 16 possible ones. But since string operators commute with the hopping terms in \eqref{Hamiltonian_B}, the corresponding three charges must have  bosonic relative statistic with respect to the moving high-energy fermions. That is, they must be `invisible' for them. This is indeed the case, since, as can be easily checked, $c$-strings move $(\bbc, \chi_\bc)$ charges. These are three fermionic charges with semionic mutual statistics.

Such an invisibility feature for several charges with nontrivial mutual statistics is not present in the Kitaev model and has a potential qualitative advantage from an experimental perspective. In this regard, in \cite{fermions_disturb} it was discussed how the high-energy fermions of the honeycomb model can spoil attempts to demonstrate the topological order. The fact that in the present model there exist topological charges invisible to high-energy fermions could simplify the operations needed to show the appearance of topological phases, because a process that only involves the invisible charges is nothing but a product of string constants of motion.


\section{Conclusions and Prospects}
\label{sect_conclusions}

We have presented a quantum lattice Hamiltonian in 2D spatial dimensions
that models many relevant and interesting features regarding systems
with quantum topological properties. An instance of this is the existence
of string-nets configurations which are constants of motion of the 
Hamiltonian and are related to fermionic processes. This is a non-perturbative result, thus valid for any
regime of coupling constants in the model.

Another prominent result is the emergence of fermionic
quasiparticles with non-trivial relative statistic, that is, anyonic ferminons. These fermions
are not free but highly interacting, which makes difficult an analytical approach.

The model we have introduced deserves further study in many directions
that we have motivated throughout our work and they are beyond the scope
of this paper. We can mention hereby several prospects for future work.

The Hamiltonian \eqref{Hamiltonian_A} has the symmetries of the lattice in Fig.~\ref{figure_lattice}(a). We could consider a more general Hamiltonian, in such a way that some of these symmetries are lost and we are left only with those of Fig.~\ref{figure_lattice}(b). This amounts to explicitly break the color symmetry of the model by substituting the three couplings $J_w$ with nine couplings $J_w^c$. The couplings $J_z^c$ correspond to the blue bonds that connect a $\bc$ and a $\bbc$ vertex, and the couplings $J_x^c$ and $J_y^c$ correspond respectively to the red and green bonds lying on $c$-plaquettes, see Fig.~\ref{figure_lattice}(b). This gives rise to a richer phase diagram and physics. For example, in the effective color code phase the differences among the $J_z^c$ couplings amounts to different gaps for $r$-, $g$- and $b$-fermions. Then different phase transitions will produce as each of the gap closes.

We have made a clear connection of the model with the topological color
code, which appears as a gapped phase. We have found perturbative indications
for the existence of other phases. Thus, it is desirable to continue
the study of the whole phase diagram of the model using other types
of techniques, including numerical simulations.

Another aspect that deserves further study is the properties of these models for 
quantum computation in a more explicit way, since here we have only focused on
their properties as far as topological order is concerned and its connections
with topological color codes.

 \noindent {\em Acknowledgements} 
 We thank X.-G. Wen and J. Vidal for useful discussions and correspondence, respectively.
We acknowledge financial support from a PFI grant of EJ-GV, DGS grants under contract
FIS2006-04885 and the ESF INSTANS 2005-10.

\end{document}